\documentclass[aps,showpacs,floatfix,11pt,preprintnumbers]{revtex4}
\usepackage{psfig}
\begin{document}
\preprint{LBNL-47452}
\title{Is the $\beta$ phase maximal ?}
\author{Javier Ferrandis}
\email{ferrandis@mac.com}
\homepage{http://homepage.mac.com/ferrandis}
\affiliation{MEC postdoctoral fellow at  the Theoretical Physics Group \\
 Lawrence Berkeley National Laboratory  \\
One Cyclotron Road, Berkeley CA 94720}
\begin{abstract}
The current experimental determination of the absolute values of the
CKM elements indicates that $2 \left| V_{ub} / V_{cb} V_{us} \right| =  (1 -z)$, with $z$ given by 
$z= 0.19 \pm 0.14$. This fact implies that  
irrespective of the form of the quark Yukawa matrices,
the measured value of the SM CP phase $\beta$ is approximately the 
maximum allowed by 
the measured absolute values of the CKM elements. This is $\beta = (\pi / 6 - z / \sqrt{3})$
for $\gamma = (\pi / 3  + z / \sqrt{3})$, which implies $\alpha=\pi/2$. 
Alternatively, assuming
that $\beta$ is exactly maximal and using the experimental measurement
$\sin(2\beta)=0.726\pm0.037$, the phase $\gamma$ is predicted to be
$\gamma= (\pi/2 - \beta) = 66.3^{\circ}\pm 1.7^{\circ}$. 
The maximality of $\beta$, if confirmed by near-future experiments,
may give us some clues as to 
the origin of CP violation. 
\end{abstract}
\maketitle
\newpage
%
\section{Introduction}
Over the past few
years, our knowledge of CP violation in particle physics
has improved significantly.
Experiments have confirmed the so-called
CKM paradigm, in that all the CP-violation processes 
measured with good precision in 
particle physics to date seem to be described by 
the CP-violating phase present in the
CKM matrix. CP-violating phases are usually parametrized in the
$(\alpha,\beta,\gamma)$ phase convention.
We have two precise B-factory measurements of $\beta$ extracted 
from the CP asymmetries in $B\rightarrow \psi K_{s}$
\cite{betaexp}:
\begin{eqnarray}
\sin(2\beta)_{\rm Belle} &=& 0.728 \pm 0.056 \pm 0.023, \\
\sin(2\beta)_{\rm BaBar} &=& 0.722 \pm 0.040 \pm 0.020.  
\label{betamea}
\end{eqnarray}
The current world average is $\sin 2\beta =0.726 \pm 0.037$,
which translates to a value for $\beta$ of, 
\begin{equation}
\beta_{\rm exp} = 23.3^{\circ} \pm 1.6^{\circ}.
\label{betaexp}
\end{equation}
The phase $\gamma$ is not so well known.
The first experimental result, which has been obtained 
at Belle, is $\gamma=(68^{+14}_{-15}\pm13\pm11)^\circ$, while  
BaBar's preliminary result is $\gamma=(88\pm41\pm19\pm10)^\circ$ 
\cite{gammaexp}. Both experiments have directly measured $\gamma$ by
studying the time-independent CP asymmetries in 
$B$ decays to $D^{0}K$ 
and $\bar{D}^{0}K$ \cite{DK}. The direct determination
carries a large combined uncertainty (statistical+systematic+model
dependent) of about $60$\% to $80\%$ \cite{gammaexp}.
There are other methods to extract $\gamma$ \cite{Zupan:2004hv}.
Recently a new method has been proposed that makes use of an 
SCET analysis of the $B\rightarrow \pi \pi$ decays.
While this method reduces the theoretical uncertainty of the final result,
the experimental error remains large \cite{Bauer:2004dg}. These measurements
are compatible with the indirect determination through CKM global fits 
\cite{fits} and earlier constraints developed in Ref.~\cite{Atwood:2001jr}, 
which provide approximately the following value for the phase $\gamma$ at $95\%$ C.L.,
\begin{equation}
\gamma_{\rm fit} \approx 61^{\circ} \pm 11^{\circ}.
\label{gammaexp}
\end{equation}
\begin{center}
\begin{figure*}    
\psfig{figure=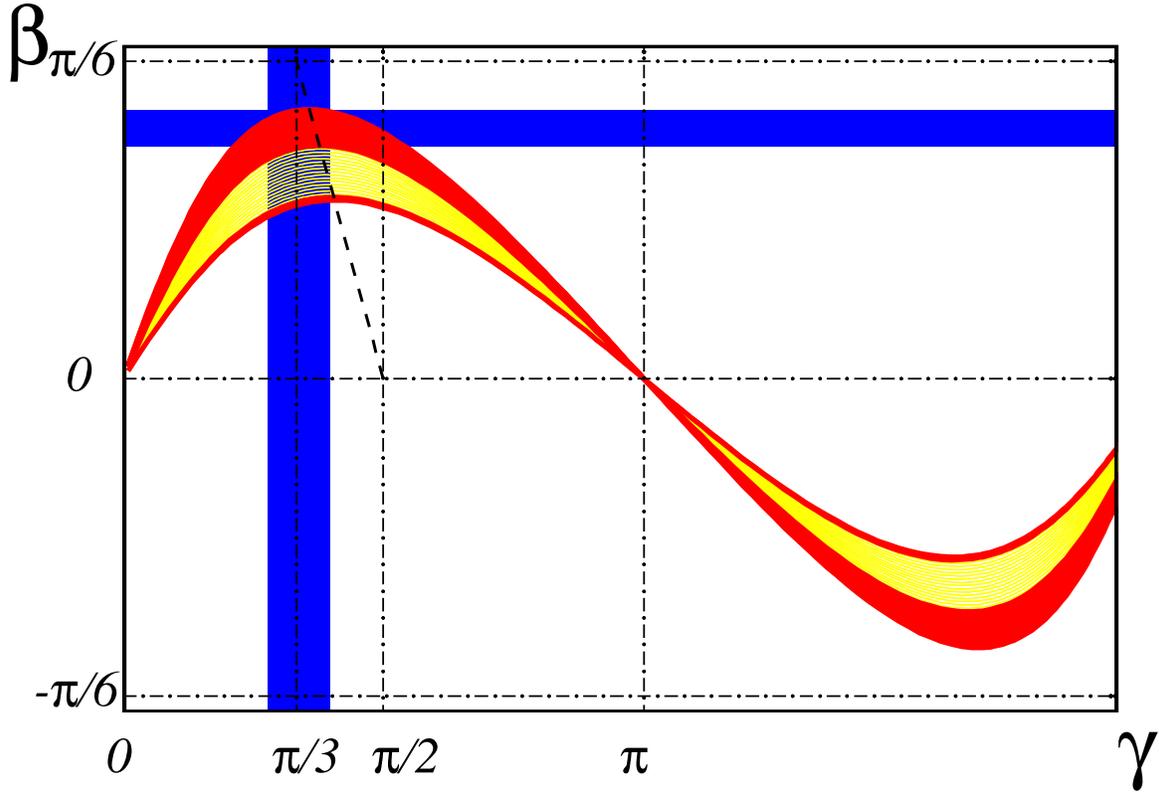,width=180mm}
\caption{\it  This figure shows the relation between the CP-violating phases $\beta$ and $\gamma$
constrained by the measured absolute values of the CKM elements. 
The blue horizontal strip 
corresponds to the measurement $\sin(2\beta)_{\rm exp}= 0.726 \pm 0.037$. The blue vertical strip 
corresponds to the 1-$\sigma$ global fit for the angle $\gamma$, 
$\gamma_{\rm fit} = 61^{\circ}\pm 11^{\circ}$. The sinusoidal red area 
corresponds to the leading-order relation between $\beta$ and $\gamma$ 
determined by unitarity (see Eq.~\ref{betagamma}). 
The sinusoidal yellow-hatched area 
corresponds to the same relation,
additionally constrained by the measurement of $\left| V_{ub} \right| / \left|V_{cb} \right|$. 
The diagonal dotted line is the line of
maximum, $\alpha=\pi/2$.}
\label{fig:betagamma}    
\end{figure*}  
\end{center}
Regarding the phase $\alpha=(\pi - \beta-\gamma)$, although none of the decays
$B\rightarrow (\pi\pi, \rho\rho,\rho\pi)$
used for the determination of $\alpha$  \cite{alphaexp,Zupan:2004hv}
gives a precise number,
the combination of the three modes \cite{Ligeti:2004ak,alphaanal} gives a result
that is in agreement with the value obtained from 
CKM global fits \cite{fits}. The latter, combining statistical and systematic uncertainities, is
approximately,
$$
\alpha_{\rm fit} \approx 100^{\circ}\pm 25^{\circ}.
$$
It is the main goal of any theory of flavor
to explain the measured CP phases, as well as the CKM
elements and fermion mass ratios.
It may be useful when 
searching for a fundamental theory of flavor 
to know {\it a priori} if there is something special about the measured values 
of $\alpha$, $\beta$ and $\gamma$
that could give us some clue regarding the origin of CP violation.
Let us suppose that we had experimental information 
on the absolute values of the CKM elements but not on the CP-violating phases. 
We could then expect that the hierarchies observed in the absolute values of the CKM elements
would constrain the range of possible values for the CP phases. 
The question is the following: are the measured values of $\beta$ and $\gamma$
a special pair out of all the possible values compatible with the 
measured absolute values of the CKM elements?
It is the main purpose of this short paper to argue that this may be the case.
We will show that the measured value of $\beta$ is approximately
the maximum allowed by the measured
absolute values of the CKM elements.
In Sec.~\ref{alter} we will introduce an alternative parametrization of the CKM matrix,
which helps to make our point more clear.
In Sec.~\ref{maxb} we will give a simple formula for the maximum value of $\beta$
and compare the numerical results with the measurements. 
In Sec.~\ref{prospects} we will briefly comment on 
near-future prospects for testing the maximality of $\beta$. 
\section{An alternative parametrization of the CKM matrix \label{alter}}
It is convenient for the subsequent discussion to 
adopt the following alternative
parametrization of the CKM matrix. The unitary matrix
${\cal V}_{\rm CKM}$ can be expressed to leading order 
as a function of three real parameters $\lambda$,$a$,$\zeta$
and a complex phase $\gamma$,
\begin{equation}
{\cal V}_{\rm CKM} =
\left[
 \begin{array}{ccc}
 1 -  \frac{\lambda^{2}}{2}  & - \lambda &   e^{ -i \gamma} \zeta  a \lambda \\
 \lambda  &  1 - \frac{\lambda^{2}}{2}  - \frac{a^{2}}{2} &   - a \\
( 1- \zeta e^{i \gamma} ) a \lambda   &  a  & 1-  \frac{a^{2}}{2} 
\end{array}
\right].
\label{CKMCP}
\end{equation}
The parameter $a$ is
given to leading order by $a ={\left| V_{cb}\right|}$.
We note that based on the experimental value for $\left| V_{cb}\right|$,
the parameter $a$ is approximately $a\approx \lambda^{2} $. Accordingly,
in the previous parametrization the 
CKM matrix is unitary to order $\lambda^{2}$, {\it i.e.} ${\cal V}_{\rm CKM}^{\dagger}
{\cal V}_{\rm CKM} = {\cal I} + {\cal O}(\lambda^{3})$. 
The parameter $\zeta$ can be determined from the absolute values of the CKM elements.
We find it convenient for the rest of the discussion 
to introduce a new parameter $ z = (1- 2 \zeta)$, which
can be expressed as,
\begin{equation}
z = 1 - \frac{ 2 \left|  V_{ub} \right| }{   \left| V_{cb}  \right| \left|V_{us} \right|  } .
\label{zeta}
\end{equation}
It is easy to obtain the relation between the new parameters $a$ and $\zeta$ and the 
parameters $A$, $\rho$ and $\eta$ of the usual Wolfenstein parametrization \cite{Wolfenstein:1983yz},
\begin{eqnarray}
A &=& a/\lambda^{2}, \quad  \zeta = ( \rho^{2} + \eta^{2})^{1/2},\\
\gamma &=& {\rm tan}^{-1} \left( \eta / \rho \right)   .
\label{Wolf}
\end{eqnarray}
The parameter $z$ can be determined from the measured absolute values of the
CKM elements. Using
the values given by the PDG collaboration, $\left|V_{us}\right| = 0.220\pm 0.0026$, 
$\left|V_{ub}\right| = 0.00367\pm 0.00047$ and $\left|V_{cb}\right| = 0.0413\pm 0.0015$ \cite{Eidelman:2004wy},
we obtain, 
\begin{equation}
z = 0.19 \pm 0.14.
\label{zexp}
\end{equation}
Using the newer and more precise value,
$\left| V_{ub}\right|/\left| V_{cb}\right| = 0.086\pm 0.008$ \cite{Battaglia:2003in},
we obtain $z = 0.22 \pm 0.08$.
\section{The maximal $\beta$ phase \label{maxb}}
It is a trivial check to prove that
the angle $\gamma$ introduced in the parametrization of
the CKM matrix given in Eq. \ref{CKMCP} coincides 
with the standard definition, 
$\gamma = {\rm Arg} \left[ - V_{ud}V^{*}_{ub} / V_{cd}V^{*}_{cb} \right]$.
Using our parametrization for the CKM matrix, 
we obtain a simple expression for the leading-order relation between the angles $\beta$ and 
$\gamma$,  
\begin{equation}
\beta
= {\rm Arg} \left[ - \frac{V_{cd}V^{*}_{cb}} {V_{td}V^{*}_{tb}} \right]
=  {\rm Arg} \left[ 1 - \zeta e^{-i \gamma}\right]
\label{betagamma}
\end{equation}
%
%
If we consider $\beta$ as a function of $\gamma$, we can determine 
the value of $\gamma$ that maximizes $\beta$, or in other words
solve the equation $d \beta /d\gamma = 0$. We find that at the maximum,
\begin{eqnarray}
\beta_{\rm max} &=& {\rm sin}^{-1} \left( \zeta \right) , \\
\gamma_{\rm max} &=& {\rm cos}^{-1} \left(\zeta \right),
\label{max}
\end{eqnarray}
where $\gamma_{\rm max}$ must be interpreted as the value of the phase $\gamma$
that maximizes $\beta$, and not the maximum value of $\gamma$.
This is equivalent to the condition, 
\begin{equation}
\beta + \gamma = \frac{\pi}{2}. 
\label{maxcond}
\end{equation}
Thus, the maximality condition predicts not $\beta$ or $\gamma$ but $\alpha$, 
$\alpha=\pi/2$. Furthermore, the measured absolute
values of the CKM elements indicate that $z\ll1$. We can therefore 
expand the maximal $\beta$ solution 
in powers of $z$ around $z=0$ and obtain the 
following expression for the maximum, 
\begin{eqnarray}
\beta^{\rm LO}_{\rm max} &=& \frac{\pi}{6} - \frac{z}{\sqrt{3}} + {\cal O}(z^{2}), \\
\gamma^{\rm LO}_{\rm max} &=& \frac{\pi}{3} + \frac{z}{\sqrt{3}} + {\cal O}(z^{2}). 
\label{betamax}
\end{eqnarray}
Using the numerical value of $z$ as calculated in Eq.~\ref{zexp}
from the absolute values of the CKM elements, 
we obtain the following numerical values for the maximum,
\begin{eqnarray}
\beta^{\rm LO}_{\rm max} &=& 23.6^{\circ} \pm 4.7^{\circ} \\
\gamma^{\rm LO}_{\rm max} &=& 66.3^{\circ} \mp 4.7^{\circ}.
\label{betatheo}
\end{eqnarray}
These values are close to the central 
measured value for $\beta$ and the global fit for $\gamma$ 
given in Eqs.~\ref{betaexp} and \ref{gammaexp} respectively.
They can be observed graphically
in figure \ref{fig:betagamma} as the intersection between
the diagonal line of maximum and the red sinusoidal strip.
If we use the value of $z$ determined using the measured 
ratio for  $\left| V_{ub}\right|/\left| V_{cb}\right| $, we obtain
a narrower strip that is shown as the hatched-yellow
area in Fig.~ \ref{fig:betagamma}.
\subsection{Next-to-leading-order corrections}
In some applications, the next-to-leading-order (NLO) corrections to the CKM
matrix play an important role. We are interested in calculating the
effect that these corrections may have on the location of the maximum
$\beta$ phase, as they may be crucial for testing the maximality of $\beta$
with precision. Requiring unitarity to order ${\cal O}(\lambda^{3})$, {\it i.e.} ${\cal V}_{\rm CKM}^{\dagger}
{\cal V}_{\rm CKM} = {\cal I} + {\cal O}(\lambda^{4})$,
we obtain, 
\begin{equation}
\left[
 \begin{array}{ccc}
 1 -  \frac{\lambda^{2}}{2}  & - \lambda (1+2a) &   e^{ -i \gamma} \zeta  a \lambda \\
 \lambda (1 +2a) &  1 - \frac{\lambda^{2}}{2}  - \frac{ a^{2}}{2} &   -a (1+\frac{a}{2} ) \\
( 1+\frac{5 a}{2} - \zeta e^{i \gamma} ) a \lambda   &  a (1+ \frac{a}{2} +\lambda^{2 }
( \zeta e^{i\gamma}-\frac{1}{2} ) ) & 1 -  \frac{a^{2}}{2} 
\end{array}
\right].
\label{CKMCPNLO}
\end{equation}
We note that the parameter $a$ is determined from $\left|V_{cb}\right|$
to NLO from $\left|V_{cb}\right|= a(1+\frac{a}{2})$.  
We can use the general definition of the $\beta$ phase given in Eq.~\ref{betagamma}
and expand in powers of $z$ and $a$. We obtain the following expression for
the maximal $\beta$ phase at NLO in the CKM matrix, 
\begin{eqnarray}
\beta^{\rm NLO}_{\rm max} &=& \frac{\pi}{6} - \frac{\eta}{\sqrt{3}} + {\cal O}(\eta^{3}), \\
\gamma^{\rm NLO}_{\rm max} &=& \frac{\pi}{3} + \frac{\eta}{\sqrt{3}} + {\cal O}(\eta^{3}), 
\label{betamaxNLO}
\end{eqnarray}
where the parameter $\eta$ is related to $z$ and $a$ by, 
\begin{equation}
\eta = (z+ 5a/2) (1 - \frac{(z+5a/2)}{6}).
\label{eta}
\end{equation}
Using the measured values of the CKM elements, we obtain $\eta= 0.28 \pm 0.14$. This translates
to the following numerical values for the maximum at NLO,
\begin{eqnarray}
\beta^{\rm NLO}_{\rm max} &=& 20.7^{\circ} \pm 4.7^{\circ}, \\
\gamma^{\rm NLO}_{\rm max} &=& 69.3^{\circ} \mp 4.7^{\circ}.
\label{betatheo}
\end{eqnarray}
Alternatively, assuming
that $\beta$ is exactly maximal and using the experimental measurement
$\sin(2\beta)=0.726\pm0.037$, the phase $\gamma$ is predicted to be
$\gamma= (\pi/2 - \beta) = 66.3^{\circ}\pm 1.7^{\circ}$.
We would like to note that the approximate maximality of
the $\beta$ phase was pointed out in the context of a particular
Yukawa ansatz in Ref.~\cite{Ferrandis:2004ti}, where the generality
of this observation was not emphasized. 
\subsection{RGE invariance}
The underlying flavor symmetry is broken
at an energy scale much higher than the electroweak scale. Therefore, 
the issue of the renormalization-scale evolution of the CP phases may be very 
relevant if $\beta$ is approximately maximal.
The evolution equations for the entries of the CKM matrix have been
known for quite a long time \cite{Ma:1979cw}.
It was pointed out \cite{Babu:1992qn} that 
using the measured hierarchies of the CKM elements
simplifies these expressions considerably. The entries $\left|{\cal V}_{ub}\right|$,
$\left|{\cal V}_{cb}\right|$, $\left|{\cal V}_{td}\right|$ and $\left|{\cal V}_{ts}\right|$
receive an identical and sizeable correction given by,
\begin{equation}
\frac{d \left| {\cal V}_{i\alpha} \right|}{d\ln \mu} = -\frac{3c}{32\pi^{2}} \left| {\cal V}_{i\alpha} \right|
(h_{t}^{2}+h_{b}^{2}).
\end{equation}
Here $c=-1$ for the SM, and $c=2/3$ for the MSSM.
This equation can be approximately solved to yield, 
\begin{equation}
\left| {\cal V}_{i\alpha}^{0} \right| \approx 
\left| {\cal V}_{i\alpha} \right|
\left( \frac{m_{W}}{\Lambda} \right)^{ \frac{3c}{32\pi^{2}} 
(h_{t}^{2}+h_{b}^{2})}.
\end{equation}
Therefore, $\left| {\cal V}_{i\alpha} \right| $ can receive a 
correction of about $+20$\% in the SM and
about $-85$\% in the MSSM.
On the other hand,
the entries $\left|{\cal V}_{us}\right|$ and
$\left|{\cal V}_{cd}\right|$ receive a tiny identical 
RGE correction given by,
\begin{equation}
\frac{d \left| {\cal V}_{us} \right|}{d\ln \mu} = -\frac{3c}{32\pi^{2}} \left| {\cal V}_{us} \right|
(h_{c}^{2} + h_{s}^{2}+ h_{t}^{2} \frac{ \left|{\cal V}_{ub}\right|^{2} -  
\left|{\cal V}_{td}\right|^{2}}{\left| {\cal V}_{cd}\right|^{2}}).
\end{equation}
This generates a correction to $\left| {\cal V}_{us}\right|$ and 
$\left|{\cal V}_{cd}\right|$, which in the SM
and MSSM is at most of the order,
\begin{equation}
\left| {\cal V}_{us} \right| \lesssim 
\left| {\cal V}_{us}^{0} \right|
\left( \frac{\Lambda}{m_{W}} \right)^{ \frac{9c}{32\pi^{2}} 
\frac{m_{s}^{2}}{m_{b}^{2}}}.
\end{equation}
This amounts to a correction of about $0.1$\%, which can be
neglected for practical purposes. Therefore
$\left| {\cal V}_{us}\right|$ and $\left|{\cal V}_{cd}\right|$ can be considered
renormalization-scale independent. As a consequence,
the parameters $\lambda$ and $\zeta$ are to a very good 
approximation renormalization-scale independent, and
the CKM matrix at the scale $\Lambda$,
${\cal V}^{0}$, adopts the same functional form as the
electroweak scale CKM matrix, with the replacement
$a \rightarrow a ( \frac{m_{W}}{\Lambda})^{ \frac{3c}{32\pi^{2}} 
(h_{t}^{2}+h_{b}^{2})}$. Since the parameter $a$ does not 
enter into Eq.~\ref{betagamma}, and the renormalization-scale 
dependence of $V_{ub}$ and $V_{cb}$ cancels in the definition of $\gamma$,
we can claim that the CP phases are to leading order 
renormalization-scale independent. Therefore, the maximality of
$\beta$ is an approximately renormalization-scale independent
fact that should be addressed by the theory of flavor, irrespective of the flavor breaking
scale.
\subsection{Maximal $\beta$ and CP-phase convention}
The usual $(\alpha,\beta,\gamma)$ phase convention
is just one particular convention out of a continuum of other possible conventions
to parametrize the CP-violating phases in the SM.
If we were using a different phase convention $(\alpha^{\prime},
\beta^{\prime},\gamma^{\prime})$, it would be a certain
combination of these phases that would become maximal, 
corresponding to the phase $\beta$.
Although this would make it 
more difficult to notice the correlation, it would have 
the same observable implications.
One may wonder whether this fact is physically relevant. 
Is it just a coincidence that one of the phases, in the 
convention that became of common use for historical reasons, is maximal?
We note that even though all phase conventions are mathematically equivalent, 
a good choice must display simple
relations to measurable quantities, and this may shed light on 
some important qualitative issues. 
For instance, in the standard convention,
$\beta$ parametrizes CPV in the mixing of the two B-meson mass eigenstates.

A related topic is the issue of maximal CP violation. 
It is known that the standard invariant measure 
of CP violation \cite{CPinvariants} is given by the Jarlskog parameter, 
${\cal J} = {\rm Im} \left[ V_{il} V_{jm} V^{*}_{im} V^{*}_{jl}\right]$ ($i\neq j$, $l \neq m$).
Using the parametrization given in Eq.~\ref{CKMCP} we obtain,
\begin{equation}
{\cal J} = \zeta a^{2} \lambda^{2} \sin(\gamma)  +{\cal O}(\lambda^{3}),
\label{Jarlskog}
\end{equation}
The Jarlskog parameter is determined experimentally to be 
${\cal J} = (3.0 \pm 0.3) \times 10^{-5}$. Substituting
$a$, $\lambda$ and $\zeta$ determined from  
the measured absolute values of the CKM
we obtain $\gamma \approx 59^{\circ}$ as the central value, which is fully consistent 
with the value provided by the current global fit.
There are several definitions of 
maximal CP violation. Originally, the term referred to the phase
that maximizes ${\cal J}$ \cite{maximalCP}.
This definition, however, is not well-defined, as 
it is phase-convention dependent. 
Moreover, it would
correspond to $\gamma=\pi/2$, which is in poor agreement with experiment,
as was pointed out a long time ago.
There have been a few proposals to define phase-convention
independent parametrizations of maximal CP violation \cite{Dunietz:1985uy},
which if true would predict an amount of CP violation 
much larger than is observed.

What is relevant is that
once the absolute values of the CKM elements have been 
fixed, the maximal $\beta$ hypothesis uniquely 
determines $\alpha$, $\beta$ and $\gamma$
, as we have shown in the previous section.
This has distinct observable implications
that are phase-convention independent.
\section{Future prospects \label{prospects}}
We have used the current measured values of the CKM elements to determine the
location of the maximal $\beta$ phase solution with about ${20\%}$ uncertainty.
This has been shown to be consistent with the current measured
value of $\beta$, which carries an uncertainty of about $5\%$.
$\left|V_{us}\right|$ and $\left|V_{cb}\right|$ are known
at the $2\%$ and $4\%$ level, respectively.
It is expected that the uncertainty in the experimental determination 
of $\left|V_{ub}\right|$ will decrease in the near future from its
present value of about $13\%$
to the present theoretical limit of $\sim 5\%$ \cite{Vub}. Such an 
improvement in the determination of $\left|V_{ub}\right|$
would allow us to calculate the maximal $\beta$ phase 
and the corresponding $\gamma$ phase with much better
precision, about $6\%$ optimistically. 
The crucial test of the maximal-$\beta$ hypothesis
will no doubt be provided 
by a precise determination of the phases $\gamma$ and $\alpha$.
The experimental situation looks favorable.
The precise measurement of $\gamma$ is the next challenge for B factories.
It is expected \cite{futuregamma,superB} that if the luminosity in the upgraded B 
factories is high enough, $\gamma$ could be determined 
with approximately $5\%$ percent uncertainty; a measurement
with an uncertainty at the level of $1\%$
will require a superB factory.
A determination of $\alpha$ at less than the $5\%$ level
may also have  to wait for the superB factories \cite{superB}. 
\section{Conclusion}
Solving a problem sometimes requires that one ask the right question.
The question of why $\beta_{\rm exp}=(23.3\pm1.6)^{\circ}$ or
the question of why $\alpha_{\rm exp} \simeq 90^{\circ}$ 
are equally puzzling.
These values of $\alpha$ and $\beta$ are at first sight two
seemingly arbitrary numbers. 
On the other hand, the observation that this 
is approximately equivalent to assuming that the phase $\beta$ is
the maximum allowed by the measured absolute values
of the CKM elements
may give us a clue about the mechanism
that determines the amount of CP violation measured in nature.
If a mechanism exists in 
the underlying theory of flavor that 
selects the maximum allowed value 
of the phase $\beta$, we can succesfully predict the measured CP violation once
the hierarchies in 
the absolute values of the CKM elements are fixed.
The maximality of $\beta$, if confirmed with precision by future 
measurements of $\gamma$ and $\alpha$, could be a guiding 
light in our search for a fundamental theory of flavor.
\acknowledgements
%
I thank Sandip Pakvasa for valuable comments. 
I also thank Hulya Guler for many suggestions. 
This work is supported by:
the Ministry of Science of Spain under grant EX2004-0238,
the D.O.E. under Contracts: DE-AC03-76SF00098 and 
DE-FG03-91ER-40676 and
by the N.S.F. under grant PHY-0098840.

\end{document}